# Electric transport properties of YBCO bicrystal films with 45° misorientation angle grown by liquid phase epitaxy

Yu. Eltsev[1], Y. Yamada[2], and K. Nakao[2]

[1]P. N. Lebedev Physical Institute, Leninskii pr. 53, Moscow, 119991, Russia

[2]Superconductivity Research Laboratory, ISTEC, 10-13, Shinonome 1-chome, Koto-ku, Tokyo, 135-0062, Japan

We report on transport properties of the grain boundaries fabricated in $YBa_2Cu_3O_{7-\delta}$ thin films grown by the liquid phase epitaxy (LPE) technique on MgO asymmetrical bicrystal substrate with 45° misorientation angle. In total around 10 samples have been studied. Substantial scatter of zero field values of the critical current density at 5K has been observed. The upper limit of $J_c \sim 10^4$ A/cm² found in our study is close to the previously reported data for 45° bicrystals grown by various physical vapour deposition methods while the minimal value of $J_c$ for the LPE grown bicrystals in striking difference to the results published before is exactly equal to zero. For samples with non-zero $J_c$ we have found a few different types of critical current dependence on magnetic field ranging from pattern reminiscent Fraunhover-like $I_c(H)$ to $I_c(H)$ profile with $I_c$ minimum at zero field.

## INTRODUCTION

Studies of the electrical transport properties of the grain boundaries (GBs) in $YBa_2Cu_3O_{7-\delta}$ (YBCO) bicrystals are important for understanding of the order-parameter symmetry of high-$T_c$ superconductors as well as for developing applications based on Josephson junctions. The GBs formed by thin films grown on bicrystal substrates by various physical vapour deposition (PVD) methods usually contain many microfacets of typical dimensions of the order of 10-100 nm that complicates interpretation of the results of transport measurements obtained on these structures [1-4]. Microstructural studies have shown that large single facet GBs (~10-30 µm) may be easily found in YBCO thin film bicrystals grown by the LPE technique [5]. Here we report on the results of measurements of transport properties of the 45$^o$ asymmetrical GBs in YBCO thin films prepared by the LPE.

## EXPERIMENTAL

YBCO films were grown by the LPE method on asymmetrical MgO bicrystal substrates with 45$^o$ misorientation angle. The typical thickness of films was about 2-3 µm. After annealing in oxygen flow at 400$^o$C for 24 hours films show superconducting transition at around 92K with $\Delta T$~1K. Bridges with widths 5-50µm were patterned across the GB. Electrical transport measurements have been performed using conventional four-probe method. Magnetic field was applied //c-axis by a solenoid coil in µ-metal shield. By sweeping magnetic field the continuous $I_c(H)$ profile was measured with positive (or negative) current-biased junction keeping voltage drop 1µV across the GB by the use of DC-picovoltmeter and simple computer program.

**RESULTS AND DISCUSSION**

In Fig. 1 we show temperature dependence of the resistance measured with current I=10μA across the GB and at H<0.05G for four samples demonstrating various types of behavior. At T~92K resistance drops in one-two orders of magnitude corresponding to the superconducting transition in the bulk of the sample and with the further decrease of temperature gradually decreases to zero. One can see strong difference in low temperature R(T) behavior reflecting the difference of $J_c$ values of these samples. For sample #1 below main superconducting transition resistance slightly increases with the decrease of temperature exhibiting no sighs of the finite value of critical current across the GB. This behavior was also found to be insensitive to the applied magnetic field. Zero $J_c$ value is in an agreement with theoretical prediction for $45^o$ asymmetrical GB with atomically flat surface [6] while non-zero value of critical current observed for samples #2 - #4 is probably due to deviations from the ideal GB in these samples.

To compare our data with the results of previous studies in Fig. 2 we show dependence of the critical current density of prepared by different PVD methods YBCO thin film asymmetrical GB junctions on the misorientation angle compiled from Ref. 1 and our data for $45^o$ bicrystals grown by the LPE technique. Similar to previous results our $J_c$ data demonstrate substantial scatter. The maximal values of the critical current density are very close for $45^o$ bicrystals grown by PVD methods as well as for the LPE bicrystals. On the other hand the minimal value of $J_c$ for the LPE grown bicrystals obtained in our study in striking contrast to previous results obtained for the GB prepared using PVD methods is exactly equal to zero. Also, $J_c$ values for several other samples are much lower in comparison with minimal $J_c$ value for PVD grown bicrystals. This observation is in a nice agreement with the results of microstructural study of the LPE grown bicrystals that undoubtedly show the better quality of the GB surface of these samples in comparison with the bicrystals prepared by various PVD methods [5].

In Fig. 3 we present magnetic field dependence of the critical current for samples #2 - #4. $I_c(H)$ curves shown in Fig. 3 demonstrate reasonable mirror symmetry with respect to H=0. Also, the results are perfectly reproducible and were not changed when we invert polarity of the DC drive across the GB. $I_c(H)$ profile for three samples looks very different ranging from Fraunhover-like $I_c(H)$ pattern for sample #4 to $I_c(H)$ dependence with $I_c$ minimum at zero field for sample#2. $I_c(H)$ profile observed for sample #3 is very similar to previously published observations for 45$^o$ asymmetrical GB in YBCO thin films prepared by PVD methods [1,4,7,8]. This behavior with complex structure of $I_c(H)$ dependence strongly deviating from conventional s-wave Fraunhofer pattern may be discussed in terms of predominant $d_{x^2-y^2}$ symmetry of the order parameter and the GB faceting. The general feature for all our $I_c(H)$ observations (including results for samples that are not shown in Fig. 3) is a very small oscillation magnetic field periodicity $\Delta H$ in comparison with a value expected from the junction size and, sometimes, $I_c(H)$ dependence reminiscent SQUID $I_c(H)$ pattern (see e.g. inset in the lower panel in Fig. 3). At present, we can only speculate on the origin of this behavior similar to Ref. 7, where SQUID-like structure was proposed due to the combination of $\pi$ and 0 junctions at the GB.

**CONCLUSIONS**

Summarizing our results we can conclude that in general transport properties of the GB fabricated in the LPE grown YBCO bicrystal films similar to those of the GB in the bicrystals prepared by various PVD methods are also affected by faceting and other possible deviations from the ideal the GB surface. On the other hand the very low $J_c$ value and, sometimes, $J_c=0$ observed for a few our samples suggests a higher quality of the GB in the selected LPE grown $45^o$ bicrystals in comparison with the bicrystals prepared by other methods. The better understanding of the relation between the transport properties of the GB fabricated by the LPE technique and its microctructure needs direct comparison of the results of studies of transport and microstructural properties of these samples.

**ACKNOWLEDGEMENTS**

This work has been supported by the New Energy and Industrial Technology Development Organization (NEDO) and partially supported by the Russian Federal Agency on science and innovation (02.513.11.3378).

**LEGENDS TO FIGURES**

**Fig. 1.**

Resistance as a function of temperature for a few samples with 45° asymmetrical GB.

**Fig. 2.**

$J_c$ dependence on misorientation angle of the GB asymmetrical junctions prepared by various PVD methods compiled from Ref. 1 and our data for several 45° bicrystals grown by the LPE method.

**Fig. 3.**

Magnetic field dependence of critical current for several 45° bicrystals grown by LPE technique. Inset in the lower panel shows $I_c(H)$ pattern in the extended magnetic field range.

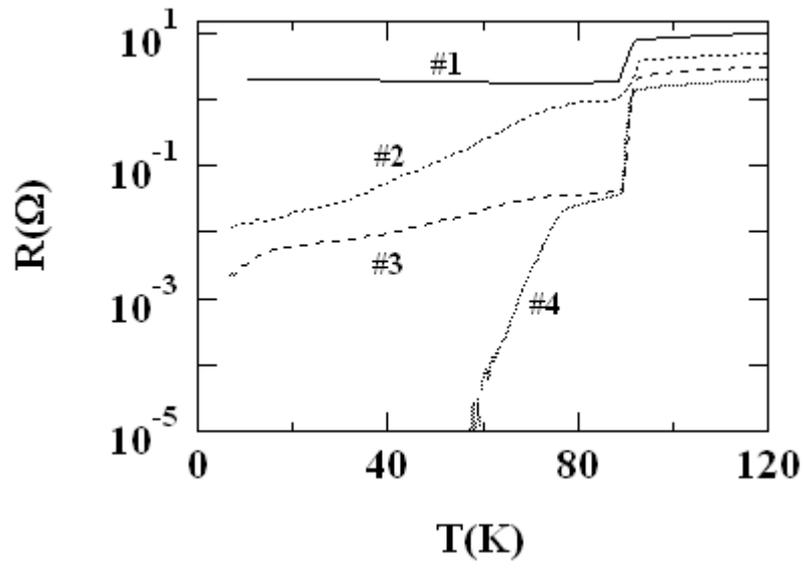

**Fig. 1**

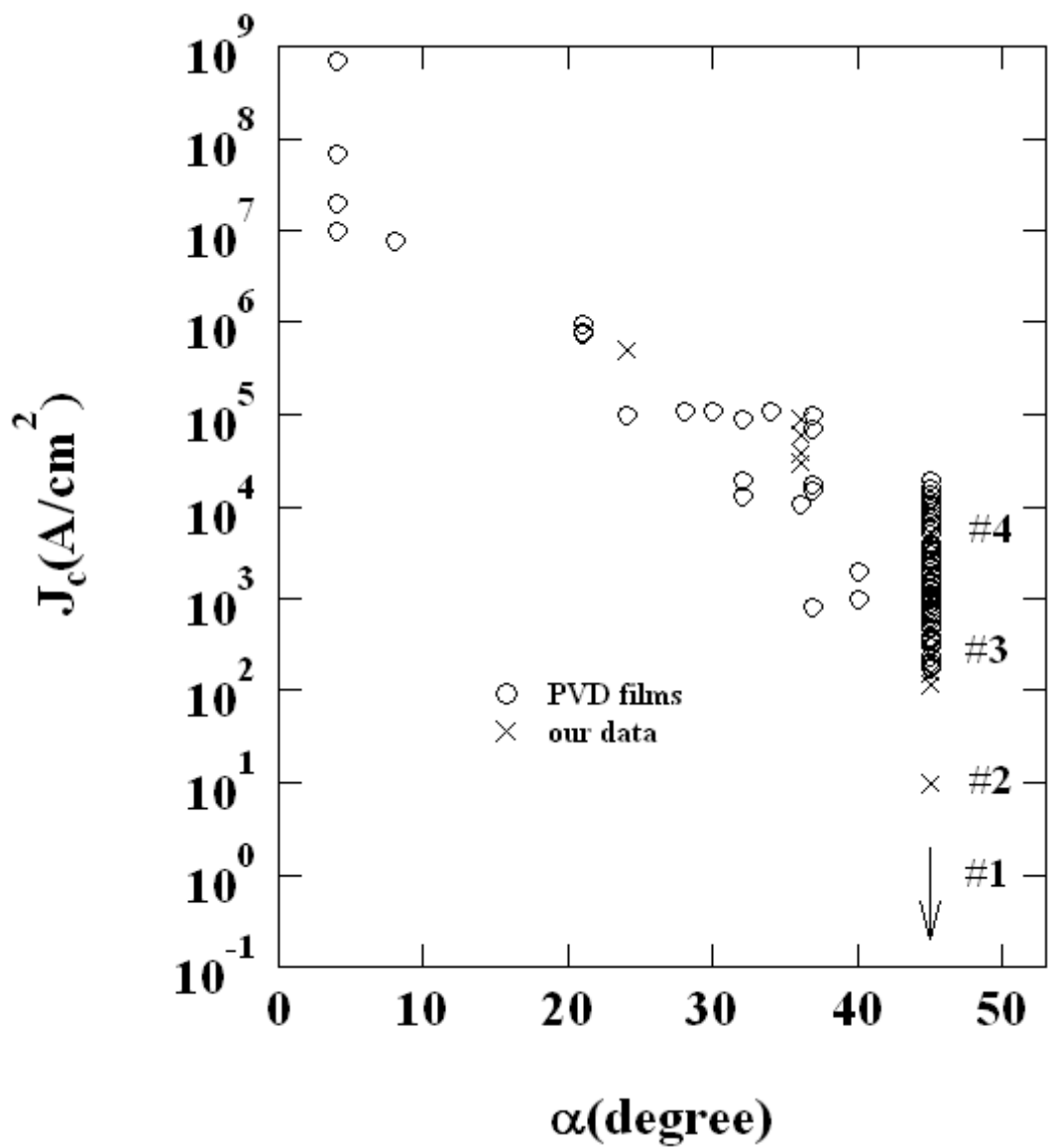

Fig. 2

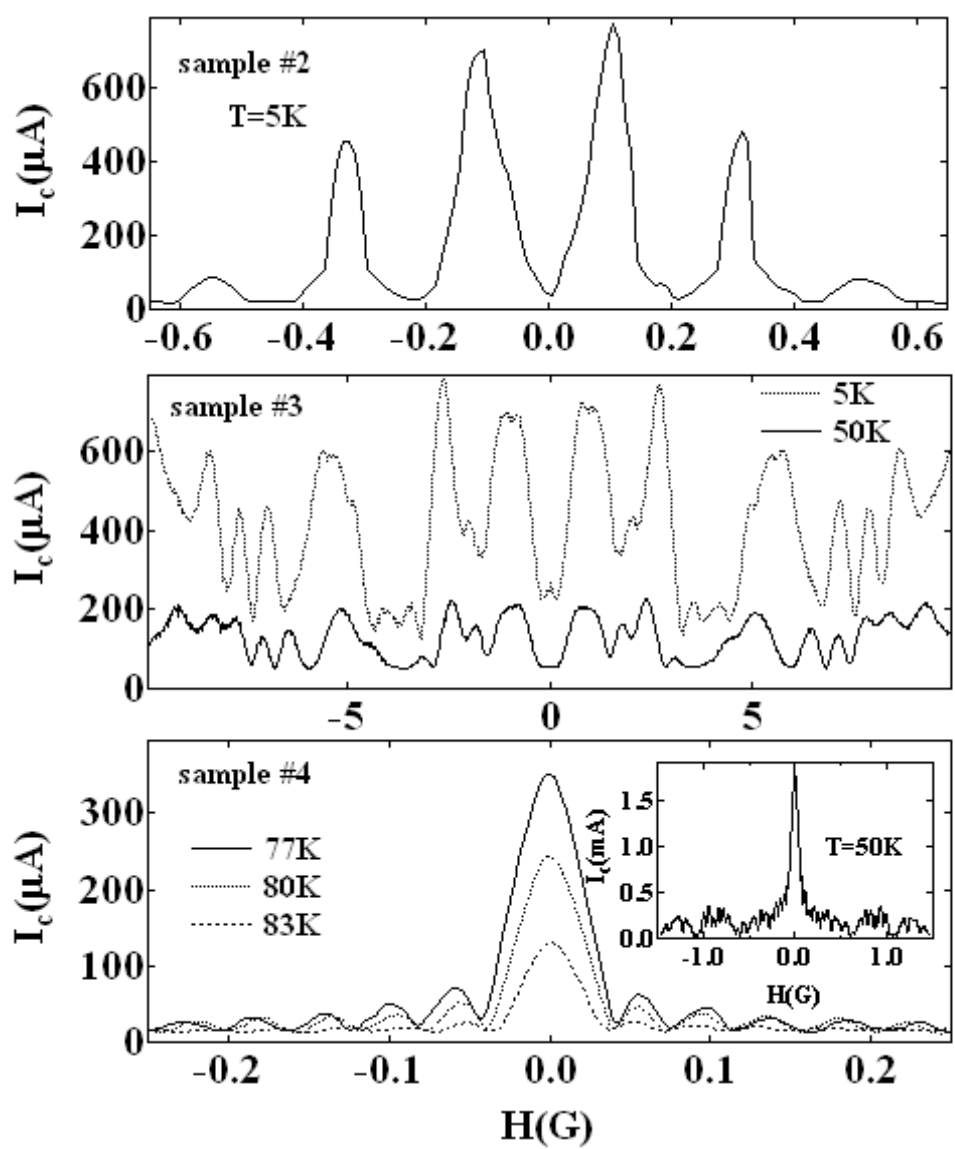

Fig. 3